\documentclass[8pt]{article}
\usepackage{spconf,amsmath,graphicx}
\usepackage{amsfonts, booktabs}
\usepackage{float}

\title{Dual-Path Cross-Modal Attention for better Audio-Visual Speech Extraction}
\name{$\text{Zhongweiyang Xu}^{*}$\text{, } $\text{Xulin Fan}^{*}$\text{, } $\text{Mark Hasegawa-Johnson}$ \thanks{*These authors contributed equally to this work}}

\address{
  University of Illinois at Urbana-Champaign}

\begin{document}
\ninept


\maketitle
\begin{abstract}
  
Audiovisual target speaker extraction is the task of separating, from an audio mixture, the speaker whose face is visible in an accompanying video.  Published approaches typically upsample the video or downsample the audio, then fuse the two streams using concatenation, multiplication, or cross-modal attention.  This paper proposes, instead, to use a dual-path attention architecture in which the audio chunk length is comparable to the duration of a video frame.  Audio is transformed by intra-chunk attention, concatenated to video features, then transformed by inter-chunk attention.  Because of residual connections, the audio and video features remain logically distinct across multiple network layers, therefore dual-path audiovisual feature fusion can be performed repeatedly across multiple layers.   When given 2-5-speaker mixtures constructed from the challenging LRS3 test set, results are about 7dB better than ConvTasNet or AV-ConvTasNet, with the performance gap widening slightly as the number of speakers increases.
  
\end{abstract}
\vspace{-5pt}

\section{Introduction}

Time-domain speech separation achieves excellent separation of multiple speech sources \cite{contasnet}.
Dual-path RNN \cite{dprnn} uses intra-chunk RNNs and inter-chunk RNNs to model short-term and long-term dependencies respectively. More recently, attention-based models are used to substitute the LSTM to get better performance \cite{sepformer, facebook1, DPTnet}. These models are now widely used as default separators for any speech enhancement  or extraction task.

One problem of these speech separation systems is that even when they're able to separate sources from mixtures, there is no clear definition of which source is the target source. Visual lip movement is able to provide such information. Humans fuse audio and video speech pre-consciously \cite{mcgurk1976hearing}, so it is natural to assume that, if a facial video is available, the visible person is the one who should be attended.

Audio-visual speech technology research includes a wide range of topics, including speech recognition, speech reconstruction, speech separation, and speech extraction. Audio-visual speech recognition (AVSR) can be performed on large in-the-wild datasets with reasonable accuracy \cite{avsr0, avsr1, avsr2, avsr3, avsr4}, thus lip movement is able to provide enough information to recognize sentences. Audiovisual feature fusion can be performed by concatenating the two feature streams at the input to a Transformer, Conformer or RNN, or by a Transformer using cross-modal attention~\cite{sterpu2018attention, AVASR, nagrani2022attention}.

Complementary to AVSR is the problem of estimating speech from silent videos.  Video-to-speech synthesis can be performed with quality that has reasonable word error rate (WER) and perceptual quality (PESQ), but that differs a lot from the original sound \cite{v2s1, v2s4}.  One  source of the difference is the lack of speaker information when only using video cues:  if speaker information is provided, the model is able to generate utterances with much better performance \cite{v2s_indivisual}.

\begin{figure*}[!h]
  \includegraphics[width=\textwidth]{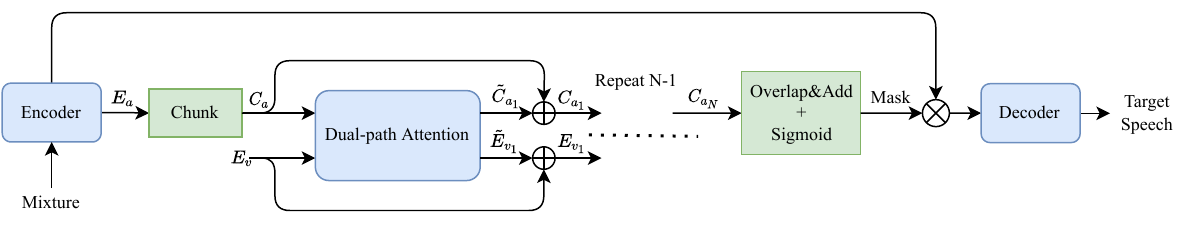}
  \vspace{-0.3in}  
  
  \caption{An overview of our mask-based audio-visual speech extraction architecture with $N$ dual-path attention modules.}
  \label{fig:mask_net}
  \vspace{-0.5cm}

\end{figure*}

Extracting speech from a mixture is a more constrained problem than video-to-speech synthesis.  Audiovisual speech extraction is typically performed using mid-level or low-level feature fusion:
typically, video features are upsampled to the audio frame rate, or vice versa, then the features are concatenated, multiplied, or combined using cross-modal attention 
\cite{conversation, concealed, tencent, multimodal_attention_speakerbeam, visualspeechenhancement, 2021_interspeech, looktolisten}. \cite{conversation, looktolisten} use concatenation fusion of frequency domain audio features and visual features to estimate separation/extraction masks on the mixture STFT. 
\cite{visualspeechenhancement} uses the same feature and fusion methods but uses an encoder-decoder structure to directly output target speech. \cite{concealed} tries to add speaker embedding into the network of \cite{conversation} to add more robustness. \cite{tencent, 2021_interspeech, multimodal_attention_speakerbeam} use a time domain separation network \cite{contasnet} 
followed by mid-level fusion of audio and video features.


This paper proposes a dual-path attention model tackling the audio-visual speech extraction problem. Dual-path attention \cite{sepformer} is  similar to the DPRNN model \cite{dprnn}, in that features are fused across time both intra-chunk and inter-chunk. 
By setting the audio chunk length comparable to the duration of a video frame, we avoid the necessity of arbitrarily upsampling the video.  Instead, each modality contributes to computations at its natural frame rate.
Each intra-chunk module uses self-attention among the audio mixture features to extract the audio's local information. In each of the inter-chunk steps, audio-visual cross-attention is used to fuse audio and visual features, while self-attention is used to combine information across chunks of time. 
Because of residual connections, this process maintains a distinction between audio and video features, hence audiovisual fusion can be performed repeatedly in multiple layers.
Our model shows superior results in terms of SI-SNR, 
PESQ, and STOI on 2-5-speaker mixtures constructed from the LRS3 dataset of real-world TED talks.

Our contributions are: 
(1) A new approach for efficient audio-visual fusion using dual-path cross attention, which solves the resolution mismatch problem between different modalities.
(2) A state-of-the-art model for audio-visual speech extraction evaluated on a challenging real-world dataset.

\vspace{-8pt}

\section{Model Architecture}

\subsection{Audio-visual Encoder}
The audio encoder is a single 1-D convolutional layer with similar structure to TasNet \cite{contasnet}, which takes raw audio with length $T$ as input and outputs an STFT-like embedding $E_{a} \in \mathbb{R}^{T'\times{D_a}}$ with feature dimension $D_a$ and time dimension $T'$. 
\vspace{-7pt}
\begin{equation}\label{eq1}
E_{a} = \text{conv1d}(E_{i})
\end{equation}
\vspace{-16pt}

On the visual side, we extract the face features $E_v \in \mathbb{R}^{T_v\times D_v}$ from video frames using a pretrained MTCNN face detection model \cite{mtcnn} and FaceNet model \cite{facenet}, where $T_v$ denotes number of frames and $D_v$ is the video feature dimension.

\vspace{-8pt}
\subsection{Masking Network}\label{sec:masking_network}
Our model is based on generating a mask on the mixture encoding. The masking network consists of a chunk operation, $N$ dual-path attention modules, an overlap and add operation to transform the chunk back to the encoding dimension, and then sigmoid activation to get the mask. The overview is shown in Figure \ref{fig:mask_net}A.

To prepare the audio embedding for the dual-path attention module, we follow the design of DPRNN \cite{dprnn} and divide $E_{a}$ into overlapping chunks with 50\% overlap. By concatenating these chunks in a new time dimension, we reshape the matrix $E_{a} \in \mathbb{R}^{T'\times D_a}$ into the 3-tensor $C_{a} \in \mathbb{R}^{S \times K \times D_a}$ where $S$ is the number of overlapping chunks and $K$ is the chunk size.

The audio 3-tensor and the visual feature matrix are then fed into $N$ connected dual-path attention modules. The dual-path attention module is covered in detail in section 2.3 below and is the main part of the masking network. Each dual-path attention module outputs two tensors which have the same dimensionality with $C_a$ and $E_v$, respectively. Residual connections are applied after each dual-path attention module. Each module's output is fed as input to the next module, until all $N$ modules are all applied. Assume $C_{a_i}$ and $E_{v_i}$ are the outputs of the $i^{\textrm{th}}$ dual-path attention module and $C_{a_0}=C_a, E_{v_0}=E_{v}$, then for $i\in[1,N]$ we have:
\begin{align}\label{cascade_net_1}
\Tilde{C}_{a_i}, \Tilde{E}_{v_i} \gets \text{DualPathAttention}(C_{a_{i-1}}, E_{v_{i-1}})\\
\label{cascade_net_2}
C_{a_i}\!=\!\text{LN}\left(\Tilde{C}_{a_i}\!+ C_{a_{i-1}}\right),\;\;E_{v_i}\!=\!\text{LN}\left(\Tilde{E}_{v_i}\!+ E_{v_{i-1}}\right)
\end{align}
After $N$ dual-path attention modules, the 50\% overlap-add operation is applied to transform the 3-tensor to a 2-D encoding with the same dimension as the audio encoder output. The 2-D encoding is passed through a sigmoid to get a mask; the Hadamard product of the mask and the audio encoder output is the encoding of the target speech.

\subsection{Dual-Path Attention}

\begin{figure}[H]
  \centering
  \includegraphics[width=\linewidth]{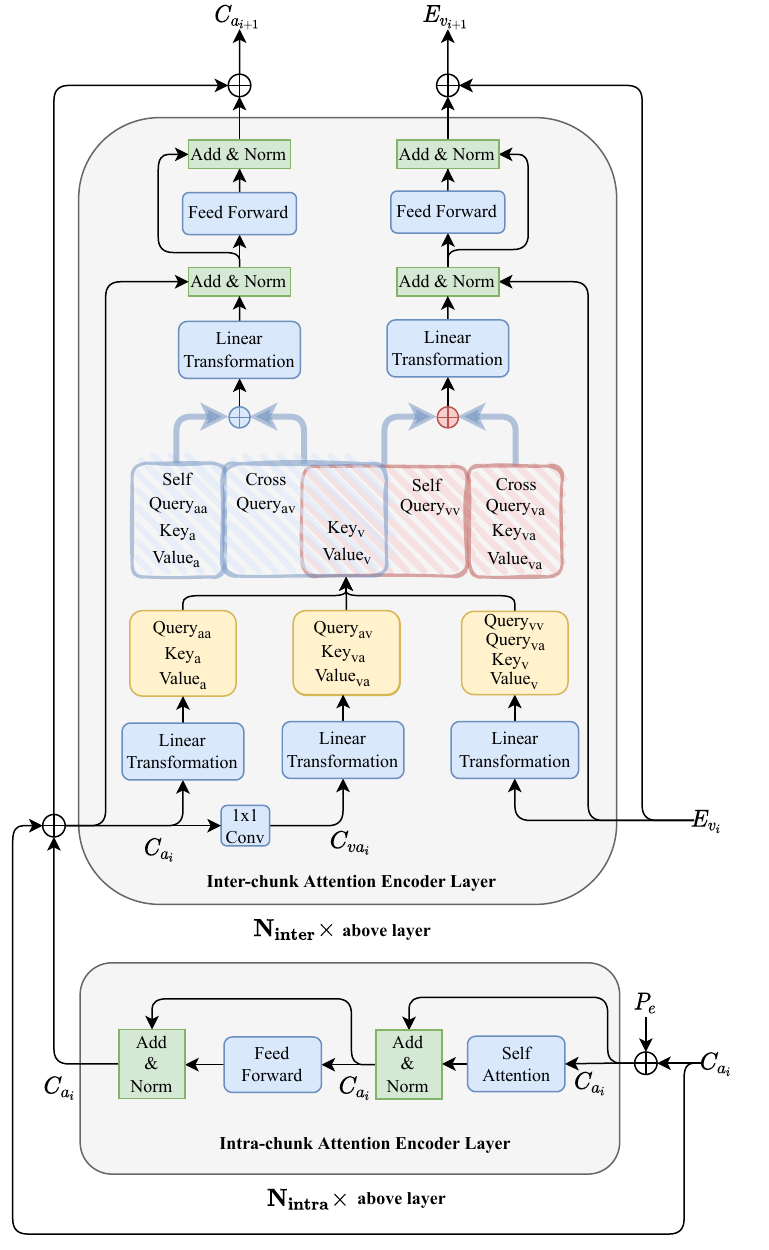}
  \vskip -0.15in

  \caption{Detailed architecture of one dual-path attention module. The module consists of $N_{intra}$ Intra-Attention encoder layers and $N_{inter}$ Inter-Attention encoder layers.}
  \label{fig:speech_production}
\vskip -0.1in
\end{figure}

Each dual-path attention module consists of $N_{\textrm{intra}}$ intra-chunk modules, followed by $N_{\textrm{inter}}$ inter-chunk attention modules, as shown in Fig.~\ref{fig:speech_production}.
Each dual-path attention module begins (at the bottom of Fig.~\ref{fig:speech_production}) by passing the audio features $C_{a_i}$ through $N_{\textrm{intra}}$ consecutive intra-chunk attention modules. Then the extracted local features are fed into the inter-chunk attention module to model long-term temporal structure and audio-visual correspondence. Residual connections after each intra-chunk and inter-chunk module improve convergence. 





The intra-chunk attention module aims to learn local temporal structure of the chunked audio feature. It consists of $N_{intra}$ layers, where each layer takes the 
chunked audio feature $C_{a} \in \mathbb{R}^{S \times K \times D_a}$ as input and outputs a tensor with the same size. Inside each layer, a MultiHeadAttention (MHA) and a position-wise feedforward layer are applied to each of the $S$ chunks separately, while residual addition and LayerNorm (LN) are applied after both MHA and FeedForward, so for all $s\in[1,S]$:
\begin{align}\label{eq3}
C_a[s,:,:] &\leftarrow \text{LN}(C_a[s,:,:] + \text{MHA}(C_a[s,:,:]))\\
C_a[s,:,:] &\leftarrow \text{LN}(C_a[s,:,:] + \text{FeedForward}(C_a[s,:,:]))
\end{align}

\vspace{-12pt}



 The inter-chunk attention module aims to deal with long-term audio-visual information. Similar to the intra-chunk attention module, each inter-chunk attention module has $N_{inter}$ encoder layers. However, different from the intra-chunk attention module, each layer in the inter-chunk attention module receives audio chunked features $C_{a} \in \mathbb{R}^{S \times K \times D_a}$ and visual features $E_{v} \in \mathbb{R}^{S \times D_v}$ as input, and then applies both self-attention for each modality and cross-attention between modalities. This makes sure that each modality can attend to both itself and the other modality, which allows both information fusion and learning long-term temporal structures.
 

Self-attention for the visual information $E_{v} \in \mathbb{R}^{S \times D_v}$ is a simple attention module as proposed in the original transformer paper \cite{Attention}. $\text{query}_{vv}$, $\text{key}_v$, and $\text{value}_v$ with dimension $\mathbb{R}^{S \times (h \times D_k)}$ are generated from linear transformations on $E_{v}$, where $S$ is the time dimension, $D_v$ is the visual feature dimension, $h$ is the number of heads, and $D_k$ is the internal dimension for each head. In the following formulas, linear transformation is abbreviated as LT.
\begin{align}\label{eq5}
\text{query}_{vv}, \text{key}_v, \text{value}_v = LT_{D_v \rightarrow (h \times D_k)}(E_v)\\
x_{vv} = \text{Attention}(\text{query}_{vv}, \text{key}_v, \text{value}_v)
\end{align}
Self-attention for the chunked audio features $C_{a} \in \mathbb{R}^{S \times K \times D_a}$ is similar to the visual side, but with an extra K dimension. We just do multihead-attention for each slice on the K dimension $C_{a_k} \in \mathbb{R}^{S\times D_a}$ and concatenate the results.
\begin{align}\label{eq5c}
\text{query}_{aa_k}, \text{key}_{a_k}, \text{value}_{a_k} = LT_{D_a \rightarrow (h \times D_k)}(C_{a}[:,k,:])\\
x_{aa}[:, k, :] = \text{Attention}(\text{query}_{aa_k}, \text{key}_{a_k}, \text{value}_{a_k})
\end{align}
For cross-attention between modalities, although the chunking step helps us match the long-term time dimension with size $S$ for the video and audio input, the audio input still has an extra short-term dimension with size $K$(chunk size) compared to the video input. We find that doing cross-attention between each slice of the audio chunked features and the video features requires a large amount of computation  and does not provide a significant performance gain. That's because for attention computation with long-term information, the attention computation is similar for each audio feature slice $C_a[:,k,:]$. So to do cross-attention efficiently, we choose to collapse the $K$ dimension of the audio features by a single $1\times1$ convolutional operation and use the collapsed audio feature $C_{va} \in \mathbb{R}^{S \times D_a}$ as the audio modality for cross-attention.
\begin{align}\label{eq5e}
        C_{va}[s,d] = 1\times1 \hspace{3px}\text{Conv}(C_{a}[s,:,d]) \\
        \text{query}_{va} = LT_{D_v \rightarrow (h \times D_k)}(E_v)\\
        \text{query}_{av}, \text{key}_{va}, \text{value}_{va} = LT_{D_a\rightarrow (h\times D_k)}(C_{va})
\end{align}


All the keys, values, and queries used in cross-attention have the same dimension $\mathbb{R}^{S \times (h \times D_k)}$, so the cross-attention computation is just a simple attention computation similar to the one in the visual information self-attention step:
\begin{align}\label{eq7}
x_{av} = \text{Attention}(\text{query}_{av}, \text{key}_v, \text{value}_v)\\
x_{va} = \text{Attention}(\text{query}_{va}, \text{key}_{va}, \text{value}_{va})
\end{align}
We add self-attention result and cross-attention result for each modality. On the visual side, $x_{vv}$ and $x_{va}$ have the same dimension $\mathbb{R}^{S \times (h \times D_k)}$, so simple element-wise addition works. On the audio side where the self-attention result $x_{aa}$ has an additional K (Chunk Size) dimension, we need to broadcast $x_{av}$ in this dimension so that the result of addition has dimension $\mathbb{R}^{S \times K \times (h \times D_k)}$. Then we use a linear transformation,
followed by LayerNorm, followed by a two-layer feedforward network (FF),
to turn the audio and video embeddings with features dimension of $(h \times D_k)$ back into $D_a$ and $D_v$ respectively:
\begin{align}\label{eq9}
\Tilde{C}_{a_i} = \text{FF}\left(\text{LN}(C_{a_{i-1}} + LT_{(h \times D_k) \rightarrow D_a}(x_{aa} + x_{av}))\right)\\
\Tilde{E}_{v_i} = \text{FF}\left(LN(E_{v_{i-1}} + LT_{(h \times D_k) \rightarrow D_v}(x_{vv} + x_{va}))\right)
\end{align}
$\Tilde{C}_{a_i}$ and $\Tilde{E}_{v_i}$ are then augmented by another residual connection and another LayerNorm, as shown in Eq.~(\ref{cascade_net_2}).

\vspace{-8pt}
\subsection{Audio Decoder}
Overlap-add with 50\% overlap transforms the chunked audio features with dimension $\mathbb{R}^{S \times K \times D_a}$ back into $\mathbb{R}^{T' \times D_a}$, which then masks the mixture encoding, as described in Section~\ref{sec:masking_network}.  Finally, the masked audio features are fed into the decoder which is a transposed convolutional layer to output the separated audio. Below $\odot$ represents Hadamard product.
\begin{align}\label{eq12}
M = \text{Sigmoid}(\text{Overlap\&Add}(C_{a_N}))\\
x_{extracted} = \text{TransposedConv1d}(M \odot E_{a})
\end{align}

\section{Experiments}
\subsection{Experimental Settings}
Hyperparameters are set as follows: $N = 3$, $N_{intra} = 4$, $N_{inter} = 4$, $K = 160$, $D_a = 256$, $D_v = 512$.  In the video processing stage, MTCNN extracts a 160$\times$160 patch for each image frame. The pretrained FaceNet model then further processes the video frames and turns each frame into a feature embedding with 512 dimensions. On the audio side, the audio encoder has a window size of 16, stride of 8, and output dimension of 256. Cropped facial regions are then fed together with audio features into the sequence of $N=3$ dual-path attention modules, each of which contains $N_{intra}=4$ intra-chunk and $N_{inter}=4$ inter-chunk attention modules. 

The configuration of multi-headed attention is the same for both intra-chunk and inter-chunk modules, each with $h = 8$ heads of internal dimension $D_k = 64$. After the attention calculation, linear layers are applied to the attention output with respectively 256 dimensions for audio and 512 dimensions for video. The FeedForward layer is just two linear layers with the intermediate feature dimension $D_{f} = 1024$. For pretraining process with the GRID dataset \cite{GRID}, the attention field covers the entire sequence. For later training and validation with the LRS3 dataset where each utterance has an 8-second duration, we apply a mask during attention calculation to limit the attention field to a 5 second duration around the current timestep.

\vspace{-8pt}
\subsection{Pretraining}

Our final evaluation of the audio-visual model is done on the LRS3 dataset \cite{LRS3}. However, we notice that LRS3 is a difficult task for the audio-visual model for three reasons. First, the video channel of LRS3 is not ideal: many image frames are non-frontal faces, and some miss the face entirely. Second, the audio channel of LRS3 has significant reverberation which makes it harder to separate out the original audio. Third, the number of speakers in LRS3 is very large, therefore it is rare for the same speaker to be repeated from one batch to the next, which makes it harder for the network to acquire useful gradients during the initial training phase. 

We therefore choose to pretrain our model on the GRID corpus \cite{GRID}, a smaller and easier dataset, before moving on to LRS3. The GRID corpus consists of 33 speakers (one speaker missing) where each speaker has 1000 3-second long utterances of simple sentences and video capturing the frontal face area. We pretrain on GRID for 10 epochs with the same training configurations mentioned below before moving to LRS3 dataset.
\vspace{-8pt}
\subsection{Training}
For the LRS3 dataset, we choose a subset of 33,000 utterances out of the whole LRS3 dataset and clip the audio and video to be 8 seconds per utterance, with a 16 kHz audio sampling rate and 25 video frames per second. From the 33000 extracted utterances, we use 27000 utterances for training, 3000 for validation, and 3000 for testing. Thus we have 60 hours of LRS3 training data, 6.67 hours for validation, and 6.67 hours for testing. The LibriSpeech and LibriMix datasets \cite{LibriSpeech, Librimix} are used as interference speech with non-overlapping train/validation/test split. For each training sample, we randomly assign 1-4 interference speakers.  The SI-SNR for the mixture is synthesized with SI-SNR (dB) drawn randomly from $\text{Uniform}(\mu-5, \mu+5)$ where $\mu$ is predefined according to the number of interference speakers. We set $\mu=0, -3.4, -5.4, -6.7$ for $1,2,3,4$ interference speakers, respectively. For the testing set, we have the exact same configuration.   

Empirically, we find that including 1-4 interfering speakers during training helps the model to improve performance on the validation set even when the validation set has only 2 speakers/mixture. Specifically, this approach helps the model to tackle difficult cases where the voice features of the target speaker and the interference speaker are similar. In these bad samples, the model is likely to make wrong choices determining the target speaker, so that even if the model is able to separate the speakers perfectly, the output is completely wrong. Including more interference speakers helps the model learn how to discriminate the target speaker from the others and thus improve the validation performance.

We use SI-SNR \cite{SISNR} as the training objective. We use Adam optimizer \cite{kingma2017adam} with initial learning rate 1e-4. We decay the learning rate by a factor of 2 when the validation SI-SNR stops improving for three consecutive epochs. The whole training process is in mixed precision with batch size 1 on one single V100 GPU. The entire training process lasts for 50 epochs.

\vspace{-2pt}
\section{Results and discussion}

We evaluate our model on our LRS3 testing set with $2,3,4,$ or $5$ speakers respectively.  Since we were unable to find published models for any previous audio-visual speaker extraction algorithm, we created audio-visual and audio-only baselines to match published near-state-of-the-art systems using ConvTasnet \cite{contasnet}.  Our audio-visual baseline is shown as AV-ConvTasnet in Table\ref{table:result}. The AV-ConvTasnet baseline uses the same 512-dimensional visual feature as our model, but its audio encoder is set to $N=512, L=32$, following the notation and recommendations in \cite{contasnet}. Then we upsample the visual features by repeating in the time dimension and concatenate them with the output of the audio encoder. The concatenated audio-visual features are then fed into the Temporal Convolutional Network with $B=256, H=1024, Sc=256, P=3, X=8, R=3$, also following the notation in \cite{contasnet}. The pre-training and training are all performed identically to our model.

\captionsetup[table]{aboveskip=-20pt}
 \begin{table}[hbt]
\caption{Speech Extraction Scores for our Model and two Baselines. SI-SNR improvement, PESQ, STOI are used as the metrics. The MIXTURE row reports the average scores of the Mixture. Note that for our model and AV-CONVTASNET, one single model is used for all the different number of speakers. For CONVTASNET, there are four models corresponding to $K=2,3,4,5$ speakers.}
\label{table:result}
\vskip 0.1in
\begin{center}
\begin{small}
\begin{sc}
\begin{tabular}{lccccr}
\toprule
Model & K & SI-SNR$i$ & PESQ & STOI \\
\midrule
Mixture    & 2& -0.05& 1.38& 0.75\\
ConvTasnet & 2& 11.23& 2.03& 0.88\\
AV-Tasnet    & 2& 11.01& 1.98& 0.88\\
Ours    & 2& 18.10& 3.08 & 0.95\\
\midrule

Mixture    & 3& -3.36& 1.12& 0.59\\
ConvTasnet & 3& 10.18& 1.38& 0.78\\
AV-Tasnet    & 3& 10.11& 1.40& 0.78\\
Ours    & 3& 17.41& 2.29 & 0.90\\
\midrule

Mixture    & 4& -5.37& 1.10 & 0.53\\
ConvTasnet & 4& 9.49 & 1.27& 0.71\\
AV-Tasnet    & 4 & 9.50 & 1.28& 0.70\\
Ours    & 4 & 17.01& 1.94 & 0.87\\
\midrule

Mixture    & 5 & -6.66& 1.09 & 0.47\\
ConvTasnet & 5 & 9.00 & 1.19& 0.65\\
AV-Tasnet    & 5 & 9.03 & 1.20& 0.65\\
Ours    & 5 & 16.60 & 1.72& 0.84\\
\bottomrule
\label{table:self-supervision}
\end{tabular}
\end{sc}
\end{small}
\end{center}
\vskip -0.4in

\end{table}
The audio-only baseline is an audio-only ConvTasnet, trained on our LRS3 training set using the permutation-invariant training (PIT) \cite{pit} SI-SNR objective. The configuration is $N=512, L=32, B=128, H=512, Sc=128, P=3, X=8, R=3$ following the notation in \cite{contasnet}. Since the conventional PIT training assumes a fixed number of sources, we train four models corresponding to $2,3,4,5$ speakers. This means the audio-only baseline models assumes a known number of speakers. The final results are summarized in Table\ref{table:result} with SI-SNRi, PESQ, STOI reported. 
As shown in Table\ref{table:result}, our model surpasses the two baseline models by a large margin. It's able to achieve 18.1dB of SI-SNR (6.87dB better than the best baseline) for the 2-speaker case where the target speech is actually recorded in the wild with reverberation. The performance gap between dual-path audiovisual attention and the best competing baseline gets slightly wider (to 7.57dB) when the number of speakers increases, thus our model has the potential to behave robustly in real-world settings.

\section{Conclusion}
In this paper, we propose a new approach to tackle modality fusion when the resolutions of different modalities vary. On top of that, we propose a speech extraction model utilizing this technique which achieves significant gain over our baseline models. 

By using the chunking process proposed by Dual-path Attention~\cite{sepformer}, we can match the $S$ (Number of chunks) dimension in the high-resolution modality with the time dimension of the low-resolution modality. We are able to match the time resolution between the audio and visual modalities naturally, thus Inter-chunk attention with modality fusion can be done readily without additional upsample operations. Utilizing this dual-path attention mechanism, our audio-visual model is able to achieve an improvement of 7dB relative to other time-domain speech extraction systems.

\bibliographystyle{IEEEtran}
\bibliography{main}

\begin{thebibliography}{10}
\providecommand{\url}[1]{#1}
\csname url@samestyle\endcsname
\providecommand{\newblock}{\relax}
\providecommand{\bibinfo}[2]{#2}
\providecommand{\BIBentrySTDinterwordspacing}{\spaceskip=0pt\relax}
\providecommand{\BIBentryALTinterwordstretchfactor}{4}
\providecommand{\BIBentryALTinterwordspacing}{\spaceskip=\fontdimen2\font plus
\BIBentryALTinterwordstretchfactor\fontdimen3\font minus
  \fontdimen4\font\relax}
\providecommand{\BIBforeignlanguage}[2]{{%
\expandafter\ifx\csname l@#1\endcsname\relax
\typeout{** WARNING: IEEEtran.bst: No hyphenation pattern has been}%
\typeout{** loaded for the language `#1'. Using the pattern for}%
\typeout{** the default language instead.}%
\else
\language=\csname l@#1\endcsname
\fi
#2}}
\providecommand{\BIBdecl}{\relax}
\BIBdecl

\bibitem{contasnet}
\BIBentryALTinterwordspacing
Y.~Luo and N.~Mesgarani, ``Conv-tasnet: Surpassing ideal time–frequency
  magnitude masking for speech separation,'' \emph{IEEE/ACM Transactions on
  Audio, Speech, and Language Processing}, vol.~27, no.~8, p. 1256–1266, Aug
  2019. [Online]. Available: \url{http://dx.doi.org/10.1109/TASLP.2019.2915167}
\BIBentrySTDinterwordspacing

\bibitem{dprnn}
Y.~Luo, Z.~Chen, and T.~Yoshioka, ``Dual-path rnn: efficient long sequence
  modeling for time-domain single-channel speech separation,'' 2020.

\bibitem{sepformer}
C.~Subakan, M.~Ravanelli, S.~Cornell, M.~Bronzi, and J.~Zhong, ``Attention is
  all you need in speech separation,'' 2021.

\bibitem{facebook1}
E.~Nachmani, Y.~Adi, and L.~Wolf, ``Voice separation with an unknown number of
  multiple speakers,'' 2020.

\bibitem{DPTnet}
J.~Chen, Q.~Mao, and D.~Liu, ``Dual-path transformer network: Direct
  context-aware modeling for end-to-end monaural speech separation,''
  \emph{arXiv preprint arXiv:2007.13975}, 2020.

\bibitem{mcgurk1976hearing}
H.~McGurk and J.~MacDonald, ``Hearing lips and seeing voices,'' \emph{Nature},
  vol. 264, no. 5588, pp. 746--748, 1976.

\bibitem{avsr0}
\BIBentryALTinterwordspacing
J.~S. Chung, A.~Senior, O.~Vinyals, and A.~Zisserman, ``Lip reading sentences
  in the wild,'' \emph{2017 IEEE Conference on Computer Vision and Pattern
  Recognition (CVPR)}, Jul 2017. [Online]. Available:
  \url{http://dx.doi.org/10.1109/CVPR.2017.367}
\BIBentrySTDinterwordspacing

\bibitem{avsr1}
\BIBentryALTinterwordspacing
T.~Afouras, J.~S. Chung, A.~Senior, O.~Vinyals, and A.~Zisserman, ``Deep
  audio-visual speech recognition,'' \emph{IEEE Transactions on Pattern
  Analysis and Machine Intelligence}, p. 1–1, 2019. [Online]. Available:
  \url{http://dx.doi.org/10.1109/TPAMI.2018.2889052}
\BIBentrySTDinterwordspacing

\bibitem{avsr2}
\BIBentryALTinterwordspacing
B.~Shi, W.-N. Hsu, K.~Lakhotia, and A.~Mohamed, ``Learning audio-visual speech
  representation by masked multimodal cluster prediction,'' in
  \emph{International Conference on Learning Representations}, 2022. [Online].
  Available: \url{https://openreview.net/forum?id=Z1Qlm11uOM}
\BIBentrySTDinterwordspacing

\bibitem{avsr3}
P.~Ma, S.~Petridis, and M.~Pantic, ``End-to-end audio-visual speech recognition
  with conformers,'' \emph{ICASSP 2021 - 2021 IEEE International Conference on
  Acoustics, Speech and Signal Processing (ICASSP)}, pp. 7613--7617, 2021.

\bibitem{avsr4}
\BIBentryALTinterwordspacing
B.~Garcia, B.~Shillingford, H.~Liao, O.~Siohan, O.~de~Pinho Forin~Braga,
  T.~Makino, and Y.~Assael, ``Recurrent neural network transducer for
  audio-visual speech recognition,'' in \emph{Proceedings of IEEE Automatic
  Speech Recognition and Understanding Workshop}, 2019. [Online]. Available:
  \url{https://arxiv.org/abs/1911.04890}
\BIBentrySTDinterwordspacing

\bibitem{sterpu2018attention}
G.~Sterpu, C.~Saam, and N.~Harte, ``Attention-based audio-visual fusion for
  robust automatic speech recognition,'' in \emph{Proceedings of the 20th ACM
  International Conference on Multimodal Interaction}, 2018, pp. 111--115.

\bibitem{AVASR}
H.~Wang, F.~Gao, Y.~Zhao, and L.~Wu, ``Wavenet with cross-attention for
  audiovisual speech recognition,'' \emph{IEEE Access}, vol.~8, pp.
  169\,160--169\,168, 2020.

\bibitem{nagrani2022attention}
A.~Nagrani, S.~Yang, A.~Arnab, A.~Jansen, C.~Schmid, and C.~Sun, ``Attention
  bottlenecks for multimodal fusion,'' 2022.

\bibitem{v2s1}
\BIBentryALTinterwordspacing
H.~Akbari, H.~Arora, L.~Cao, and N.~Mesgarani, ``Lip2audspec: Speech
  reconstruction from silent lip movements video,'' \emph{CoRR}, vol.
  abs/1710.09798, 2017. [Online]. Available:
  \url{http://arxiv.org/abs/1710.09798}
\BIBentrySTDinterwordspacing

\bibitem{v2s4}
A.~Ephrat and S.~Peleg, ``Vid2speech: speech reconstruction from silent
  video,'' in \emph{2017 IEEE International Conference on Acoustics, Speech and
  Signal Processing (ICASSP)}.\hskip 1em plus 0.5em minus 0.4em\relax IEEE,
  2017.

\bibitem{v2s_indivisual}
\BIBentryALTinterwordspacing
K.~R. Prajwal, R.~Mukhopadhyay, V.~Namboodiri, and C.~V. Jawahar, ``Learning
  individual speaking styles for accurate lip to speech synthesis,''
  \emph{CoRR}, vol. abs/2005.08209, 2020. [Online]. Available:
  \url{https://arxiv.org/abs/2005.08209}
\BIBentrySTDinterwordspacing

\bibitem{conversation}
T.~Afouras, J.~S. Chung, and A.~Zisserman, ``The conversation: Deep
  audio-visual speech enhancement,'' 2018.

\bibitem{concealed}
------, ``My lips are concealed: Audio-visual speech enhancement through
  obstructions,'' 2019.

\bibitem{tencent}
J.~Wu, Y.~Xu, S.-X. Zhang, L.-W. Chen, M.~Yu, L.~Xie, and D.~Yu, ``Time domain
  audio visual speech separation,'' 2019.

\bibitem{multimodal_attention_speakerbeam}
H.~Sato, T.~Ochiai, K.~Kinoshita, M.~Delcroix, T.~Nakatani, and S.~Araki,
  ``Multimodal attention fusion for target speaker extraction,'' 2021.

\bibitem{visualspeechenhancement}
A.~Gabbay, A.~Shamir, and S.~Peleg, ``Visual speech enhancement,'' 2018.

\bibitem{2021_interspeech}
Y.~Luo, J.~Wang, L.~Xu, and L.~Yang, ``{Multi-Stream Gated and Pyramidal
  Temporal Convolutional Neural Networks for Audio-Visual Speech Separation in
  Multi-Talker Environments},'' in \emph{Proc. Interspeech 2021}, 2021, pp.
  1104--1108.

\bibitem{looktolisten}
\BIBentryALTinterwordspacing
A.~Ephrat, I.~Mosseri, O.~Lang, T.~Dekel, K.~Wilson, A.~Hassidim, W.~T.
  Freeman, and M.~Rubinstein, ``Looking to listen at the cocktail party,''
  \emph{ACM Transactions on Graphics}, vol.~37, no.~4, p. 1–11, Aug 2018.
  [Online]. Available: \url{http://dx.doi.org/10.1145/3197517.3201357}
\BIBentrySTDinterwordspacing

\bibitem{mtcnn}
\BIBentryALTinterwordspacing
K.~Zhang, Z.~Zhang, Z.~Li, and Y.~Qiao, ``Joint face detection and alignment
  using multitask cascaded convolutional networks,'' \emph{IEEE Signal
  Processing Letters}, vol.~23, no.~10, p. 1499–1503, Oct 2016. [Online].
  Available: \url{http://dx.doi.org/10.1109/LSP.2016.2603342}
\BIBentrySTDinterwordspacing

\bibitem{facenet}
\BIBentryALTinterwordspacing
F.~Schroff, D.~Kalenichenko, and J.~Philbin, ``Facenet: A unified embedding for
  face recognition and clustering,'' \emph{2015 IEEE Conference on Computer
  Vision and Pattern Recognition (CVPR)}, Jun 2015. [Online]. Available:
  \url{http://dx.doi.org/10.1109/CVPR.2015.7298682}
\BIBentrySTDinterwordspacing

\bibitem{Attention}
\BIBentryALTinterwordspacing
A.~Vaswani, N.~Shazeer, N.~Parmar, J.~Uszkoreit, L.~Jones, A.~N. Gomez,
  L.~Kaiser, and I.~Polosukhin, ``Attention is all you need,'' \emph{CoRR},
  vol. abs/1706.03762, 2017. [Online]. Available:
  \url{http://arxiv.org/abs/1706.03762}
\BIBentrySTDinterwordspacing

\bibitem{GRID}
\BIBentryALTinterwordspacing
M.~Cooke, J.~Barker, S.~Cunningham, and X.~Shao, ``An audio-visual corpus for
  speech perception and automatic speech recognition,'' \emph{The Journal of
  the Acoustical Society of America}, vol. 120, no.~5, pp. 2421--2424, 2006.
  [Online]. Available: \url{https://doi.org/10.1121/1.2229005}
\BIBentrySTDinterwordspacing

\bibitem{LRS3}
T.~Afouras, J.~S. Chung, and A.~Zisserman, ``Lrs3-ted: a large-scale dataset
  for visual speech recognition,'' in \emph{arXiv preprint arXiv:1809.00496},
  2018.

\bibitem{LibriSpeech}
V.~Panayotov, G.~Chen, D.~Povey, and S.~Khudanpur, ``Librispeech: An asr corpus
  based on public domain audio books,'' in \emph{2015 IEEE International
  Conference on Acoustics, Speech and Signal Processing (ICASSP)}, 2015, pp.
  5206--5210.

\bibitem{Librimix}
J.~Cosentino, M.~Pariente, S.~Cornell, A.~Deleforge, and E.~Vincent,
  ``Librimix: An open-source dataset for generalizable speech separation,''
  \emph{arXiv preprint arXiv:2005.11262}, 2020.

\bibitem{SISNR}
J.~Le~Roux, S.~Wisdom, H.~Erdogan, and J.~R. Hershey, ``Sdr--half-baked or well
  done?'' in \emph{ICASSP 2019-2019 IEEE International Conference on Acoustics,
  Speech and Signal Processing (ICASSP)}.\hskip 1em plus 0.5em minus
  0.4em\relax IEEE, 2019, pp. 626--630.

\bibitem{kingma2017adam}
D.~P. Kingma and J.~Ba, ``Adam: A method for stochastic optimization,'' 2017.

\bibitem{pit}
M.~Kolbæk, D.~Yu, Z.-H. Tan, and J.~Jensen, ``Multi-talker speech separation
  with utterance-level permutation invariant training of deep recurrent neural
  networks,'' 2017.

\end{thebibliography}


\end{document}